\documentclass[%
 reprint,
superscriptaddress,
 amsmath,amssymb,
 aps,
]{revtex4-2}
\usepackage{graphicx}
\usepackage{dcolumn}
\usepackage{bm}
\usepackage{wrapfig}
\usepackage{mhchem}
\usepackage{amssymb}
\usepackage{amsmath}
\usepackage{supertabular}
\usepackage[table,xcdraw]{xcolor}
\begin{document}

\title{Beyond Lithium-Ion Batteries: Are Effective Electrodes Possible for Alkaline and Other Alkali Elements? Exploring Ion Intercalation in Surface-Modified Few-Layer Graphene and Examining Layer Quantity and Stages}

\author{Yu-Hsiu Lin}
 \affiliation{Department of Chemical Engineering \& Materials Science, Michigan State University, East Lansing, Michigan 48824, United States.}
\author{Jose L. Mendoza-Cortes}
 \email{jmendoza@msu.edu}
\affiliation{Department of Physics \& Astronomy, Michigan State University, East Lansing, Michigan 48824, United States.}
\affiliation{Department of Chemical Engineering \& Materials Science, Michigan State University, East Lansing, Michigan 48824, United States.}

\date{\today}

\begin{abstract}

In the pursuit of reliable energy storage solutions, the significance of engineering electrodes cannot be overstated. Previous research has explored the use of surface modifiers (SMs), such as single-side fluorinated graphene, to enhance the thermodynamic stability of ion intercalation when applied atop few-layer graphene (FLG). As we seek alternatives to lithium-ion batteries (LIBs), earth-abundant elements like sodium and potassium have emerged as promising candidates. However, a comprehensive investigation into staging intercalation has been lacking thus far. By delving into staging assemblies, we have uncovered a previously unknown intercalation site that offers the most energetically favorable binding. Here we study the first three elements in both alkali (Li, Na, K) and alkaline (Be, Mg, Ca) earth metals. Furthermore, the precise mechanism underlying this intercalation system has remained elusive in prior studies. In our work, we employed Density Functional Theory (DFT) calculations with advanced hybrid functionals to determine the electrical properties at various stages of intercalation. This approach has been proven to yield more accurate and reliable electrical information. Through the analysis of projecting density of states (PDOS) and Mulliken population, we have gained valuable insights into the intricate interactions among the SM, ions, and FLG as the ions progressively insert into the structures. Notably, we expanded our investigation beyond lithium and explored the effectiveness of the SM on ions with varying radii and valence, encompassing six alkali and alkaline earth metals. Additionally, we discovered that the number of graphene layers significantly influences the binding energy. Our findings present groundbreaking concepts for material design, offering diverse and economically viable alternatives to LIBs. Furthermore, they serve as a valuable reference for fine-tuning electrical properties through staging intercalation and the application of SMs.

\end{abstract}

\maketitle

\section{Introduction}

The usage of lithium-ion batteries (LIBs) has become omnipresent in our daily lives, leading to a growing concern about potential supply shortages of lithium \cite{gruber2011global,olivetti2017lithium,scrosati2000recent,pramudita2017initial}. Consequently, there is a significant focus on developing batteries that are both more efficient and cost-effective \cite{tarascon2001issues,dunn2011electrical,armand2008building,goodenough2010challenges,choi2016promise}. One way to address these concerns is by effectively utilizing lithium, such as by improving its electric capacity through engineering the electrode materials. Carbon-based materials like graphite and few-layer graphene have shown promise in enhancing ion intercalation \cite{yabuuchi2014research,lotfabad2014high,goodenough2013li,li2019intercalation,brandt1994historical}. In addition to optimizing lithium usage, researchers are exploring alternative intercalants to lithium, such as sodium and potassium atoms, due to their similar electronic properties \cite{palomares2012ion,zhang2019approaching,kundu2015emerging,slater2013sodium,kim2018recent}. These alternatives are being investigated as promising options for intercalation within layer-structured electrodes. Furthermore, some alternative metals offer more competitive prices compared to lithium. For example, sodium and magnesium can be economical alternatives based on their carbonate counterparts \cite{masse2018beyond}. However, these alternative intercalants present certain challenges that differ from those of lithium. These challenges can adversely affect the intercalation process and result in reduced storage capacity \cite{jian2015carbon,liu2016origin,wen2014expanded}. One disadvantage is their larger atomic radius compared to lithium, which introduces an additional energy barrier for intercalation and requires a greater degree of geometric transformation to accommodate more spacious host sites \cite{li2019intercalation,nobuhara2013first}.

\begin{figure*}
\includegraphics[width=\textwidth]{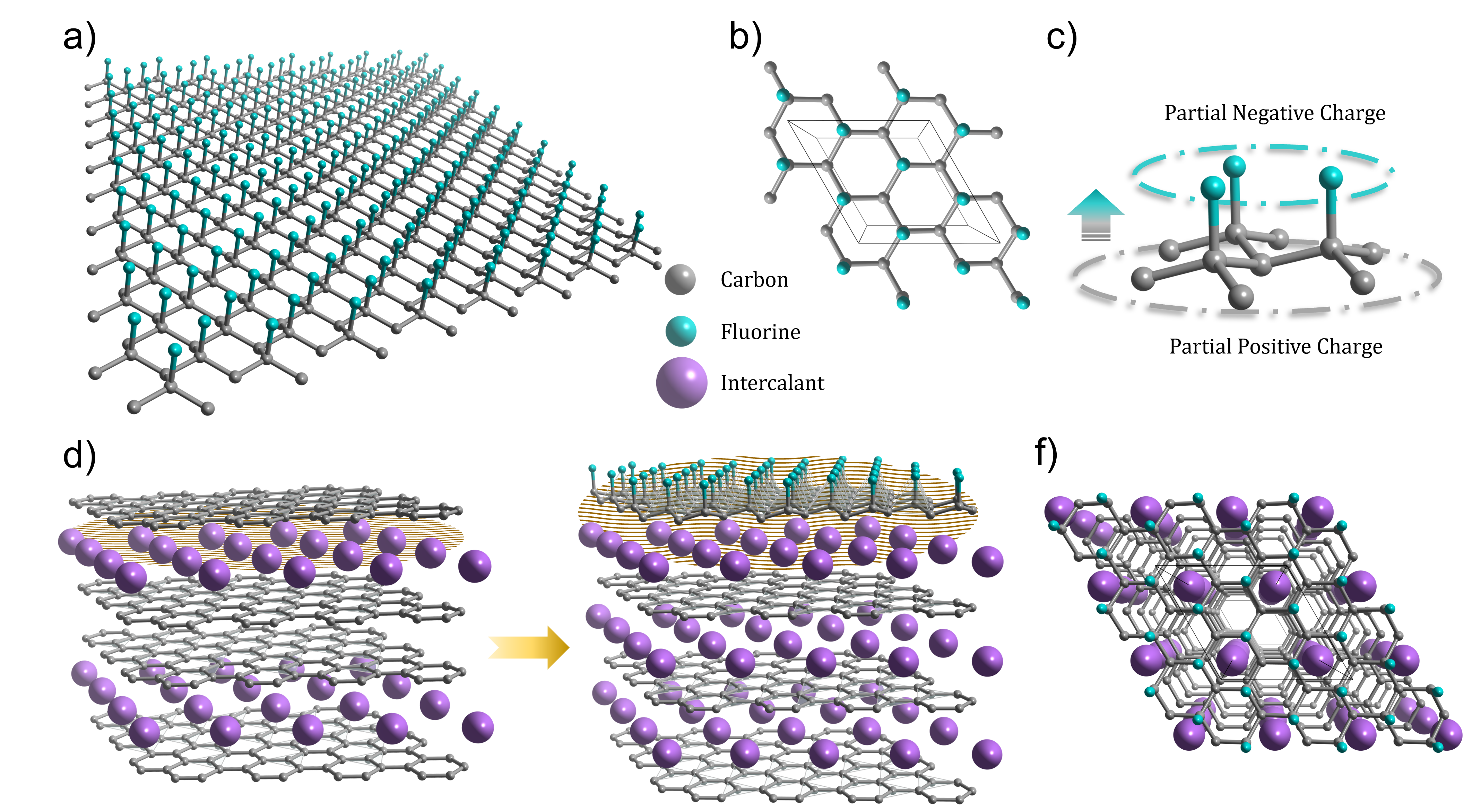}
\centering
\caption{\footnotesize Key principles explored in this study: (a) The side view and (b) top view of a Two-Dimensional 50\% Fluorinated Monolayer Graphene (2D FSM) Layer; (c) The side view of an FSM layer showing its intrinsic dipole moment; (d) Augmented Intercalant Capacity via FSM Insertion, highlighting electron transfer pathways (shown in the yellow background) facilitating interaction among various components within the system; (f) The top view of fluorinated 4LG with intercalants at stage 1.}
\label{fig:Scheme}
\end{figure*}

Hence, exploring potential alternatives with comparable atomic radii provides a new avenue for research. One such alternative is the magnesium atom, which possesses a slightly smaller radius than lithium and can also act as a multivalent ion, thereby enhancing battery capacity by carrying more charge per intercalant. In this study, both electrode engineering and lithium alternatives were investigated. For the electrodes, the focus was on a fluorinated surface modifier (FSM) applied to the top of few-layer graphene (FLG). While graphenes can serve as electrodes and have demonstrated higher lithium storage capacity than graphite \cite{yoo2008large,brownson2011overview,wang2009electrochemical,lian2010large,kuhne2017ultrafast,hui2016layer}, sodium (Na) intercalation remains unfavorable \cite{ling2014boron,moriwake2017sodium}. Therefore, various modifications of graphenes, such as defects and surface modifiers, were employed to enhance intercalation and ion transport \cite{gong2017boron,share2016role,olsson2021investigating,ferrari2013raman,ghaderi2010first,ghimbeu2018insights}. One example is fluorine-doped FLG, which capitalizes on the high electronegativity of fluorine to improve capacity \cite{ju2016direct}. Moreover, fluorinated graphenes can be easily synthesized and possess unique electronic properties \cite{jeon2011fluorographene,withers2011nanopatterning,csahin2011structures}. These considerations led to the development of FLG with FSM. Building upon our previous work \cite{nijamudheen2020impact}, a monolayer of fluorinated graphene on top of four-layer graphene (4LG) has been shown theoretically to enhance the intercalation of lithium, sodium, and potassium atoms within the graphene layers, exhibiting lower binding energies compared to pristine 4LG. This two-dimensional modification not only reinforces lithium intercalation in FLG but also makes the substitution of lithium with other atoms more promising.
 
However, the precise mechanism by which the FSM influences the intercalation capability in FLG remains unclear and necessitates further investigation. Considering the high electronegativity of fluorine atoms, we propose that the electric potential difference induced by the FSM acts as both the driving force for charge transfer and the stabilizer for the intercalant. To verify this assumption, we computed the electronic properties (e.g., PDOS, Mulliken analysis) of ion-intercalated FLG with and without the FSM, in addition to the relative binding energy. For this purpose, we employed density functional theory (DFT) with the advanced hybrid functional B3LYP-D3, which enables accurate predictions of electronic properties, albeit at a higher computational cost compared to the previously used PBE functional \cite{mardirossian2017thirty}. Moreover, the intercalation mechanism of lithium in carbon-based layers is known to involve a gradual staging process, resulting in ordered intercalation at varying concentrations \cite{levi1997mechanism,hui2021nernstian,dahn1991phase}. Similarly, potassium (K) and sodium (Na) ion intercalation can be described as staging procedures as well \cite{jian2015carbon,kim2015sodium,hui2018achieving}. However, FLG with the same ion concentration can exhibit different assemblies, with ions intercalated within different layers. By introducing a single-side fluorinated SM, the number of potential assemblies is further increased due to the breaking of structural symmetry. Unfortunately, limited research has been conducted on intercalant assemblies in FLG. We believe that these assemblies can impact thermodynamic stability and electronic properties. In this study, we not only discuss different staging structures but also consider how they host intercalants. Interestingly, staging structures with intercalation sites between the FSM and its neighboring FLG layer are more likely to be energetically favorable assemblies. This finding underscores the significance of the FSM, as it not only reduces binding energies but also creates more stable host sites.

To extend the scope of the FSM's applicability, we proceeded to study the intercalation of the first three elements in both alkali and alkaline earth metals within FLG. Our objective was to unravel the interactions between the FSM and ions with varying atomic radii and multivalency. The energetic outcomes revealed that, irrespective of the type of intercalants, all scenarios involving the FSM exhibited enhanced binding between the intercalants and the graphene substrate. However, their respective electronic interactions displayed distinct characteristics. Additionally, the number of graphene layers emerged as a critical factor affecting the binding energy. Fewer layers in FLG corresponded to lower binding energies, underscoring the significance of FLG exfoliation from graphite when considering electrode materials. In summary, these findings present multiple potential alternatives to LIBs and shed light on the long-standing ambiguity surrounding surface modifiers and staging intercalation. Furthermore, the implications of this research extend beyond the realm of batteries, as it offers valuable insights for tuning the stability and electrical properties of layered materials through the application of SMs or by varying the density of intercalants and the number of layers.

\begin{figure}[h]
    \centering
    \includegraphics[width=0.47\textwidth]{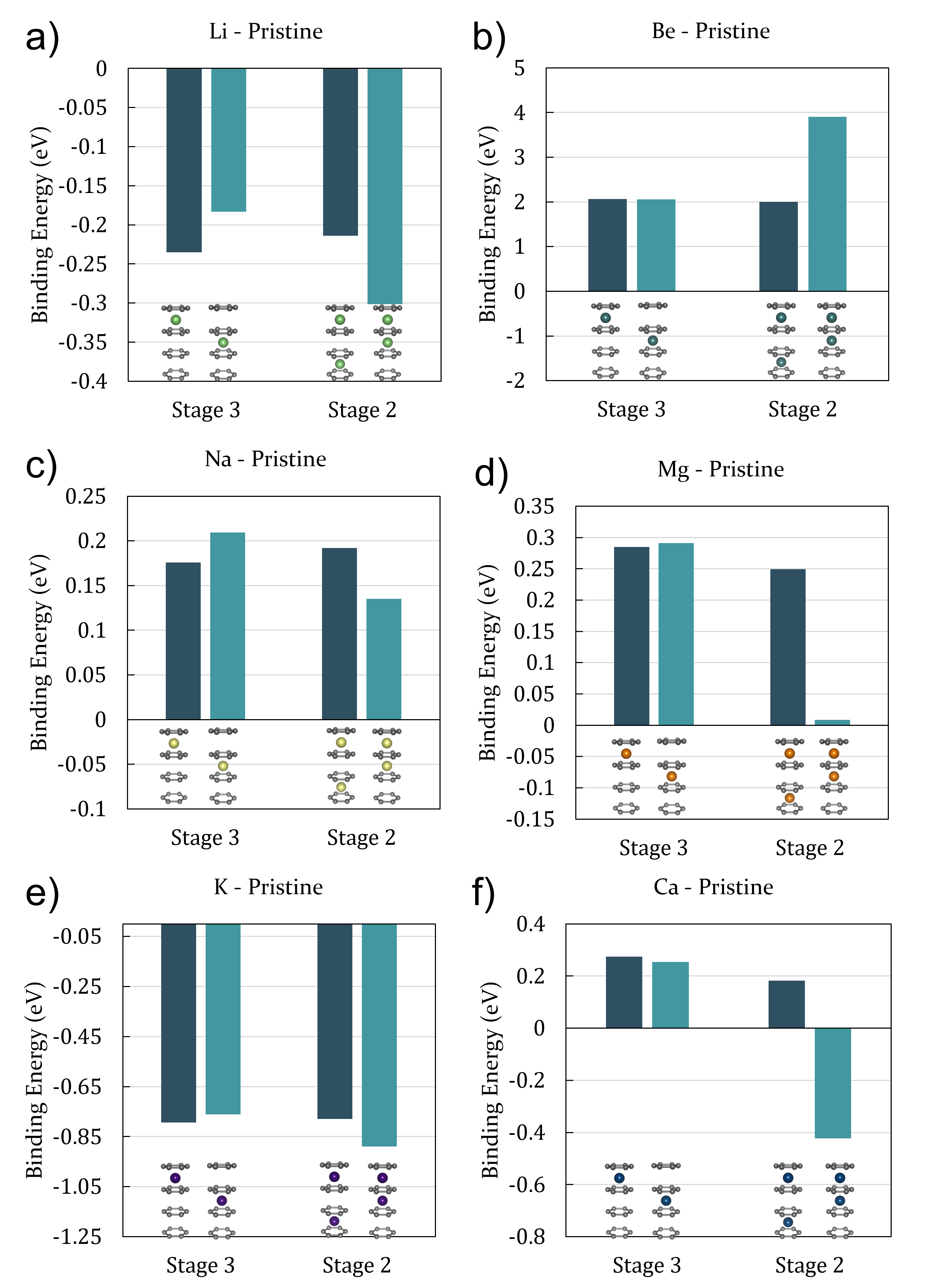}
\caption{\footnotesize Comparison of binding energies between staging assembles at stage 3 and stage 2, involving ion intercalations of (a) Li, (b) Be, (c) Na, (d) Mg, (e) K, and (f) Ca within pristine 4LG. The schematic diagrams, inserted for clarity, correspond to the respective binding energies.}
    \label{fig:BE_pristine}
\end{figure}

\begin{figure}[h]
    \centering
    \includegraphics[width=0.47\textwidth]{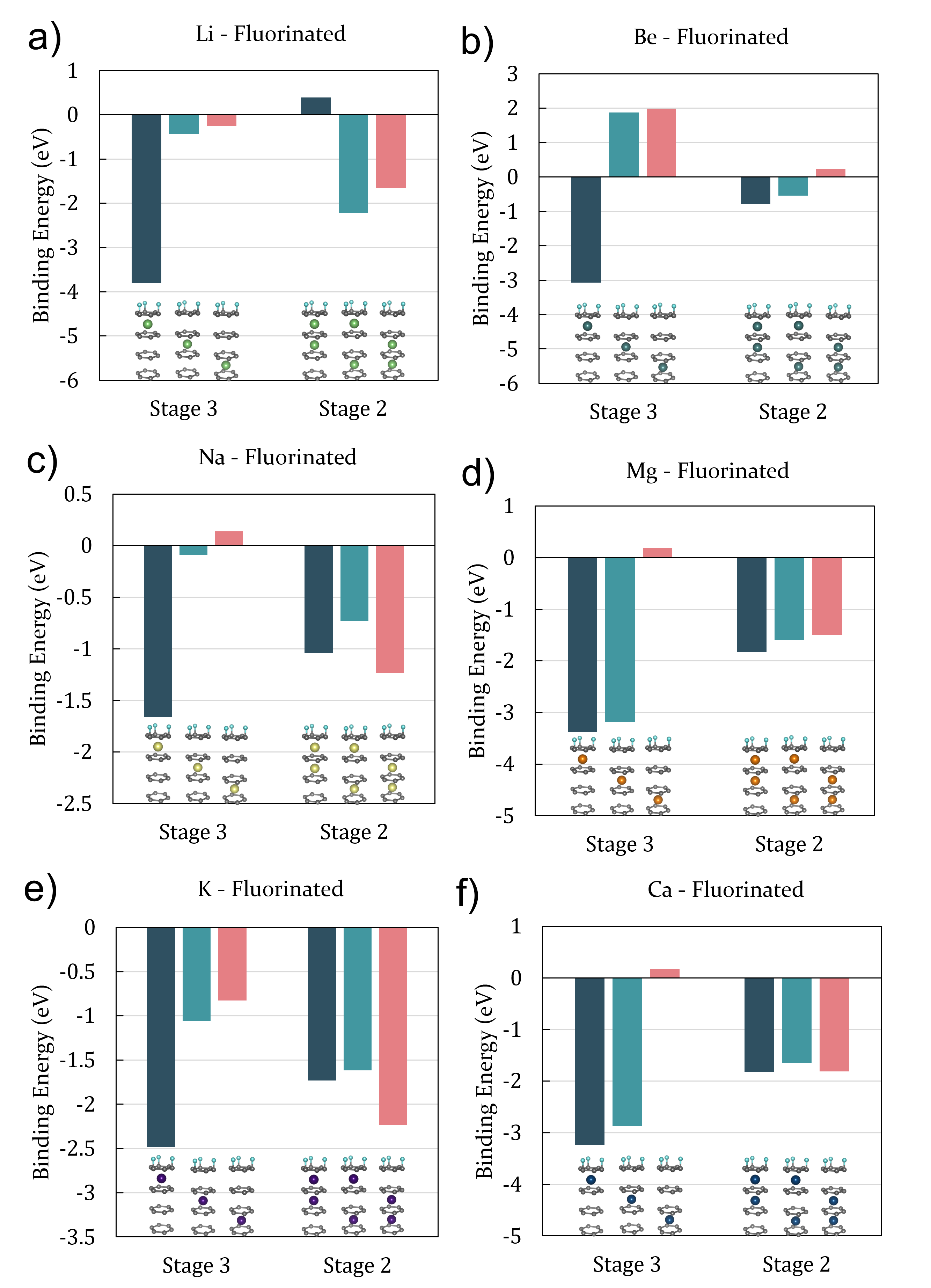}
\caption{\footnotesize Comparison of binding energies for staging assembles at stage 3 and stage 2, involving ion intercalations of (a) Li, (b) Be, (c) Na, (d) Mg, (e) K, and (f) Ca within fluorinated 4LG. The schematic diagrams, inserted for clarity, correspond to the respective binding energies.}    
    \label{fig:BE_fluorinated}
\end{figure}

\section{Methodology}

Calculations were carried out using unrestricted periodic DFT with the B3LYP hybrid functional \cite{becke1992density,lee1988development}, as implemented in the CRYSTAL17 code \cite{dovesi2018quantum} with Gaussian-type basis sets. Since the slab structures with multiple layers are represented within extended unit cells, the D3 dispersion correction \cite{grimme2006semiempirical,grimme2010consistent,grimme2011effect}, accounting for long-range Coulomb correlation and weak London forces, was combined with the B3LYP functional. Specifically, Gaussian triple-zeta valence with polarization (TZVP) basis sets \cite{peintinger2013consistent} was used to represent the orbitals of each element, and full optimization was performed to adjust both atom positions and unit cell parameters. To prevent interactions between cells and simulate surface conditions, a ${\sim}$500 Å vacuum layer along the z-axis was employed for the slab structures.
For self-consistent field (SCF) calculations, direct inversion of the iterative subspace (DIIS) \cite{pulay1980convergence,pulay1982improved} was utilized to achieve an energy convergence criterion of 2.72 $\times$ 10$^{–6}$ eV. The convergence criteria for geometry optimization were set to an RMS force of 1.54 $\times$ 10$^{–2}$ eV/{\AA}, a max force of 2.31 $\times$ 10$^{–2}$ eV/{\AA}, an RMS displacement of 6.35 $\times$ 10$^{–4}$ {\AA}, and a max displacement of 9.53 $\times$ 10$^{–4}$ {\AA}. Subsequently, based on the optimized structures, relative electronic properties, including density of states (DOS) and Mulliken analysis, were calculated. To obtain higher-resolution electronic results, the first Brillouin zone (${\sim}$$2\pi$ $\times$ $1/60$ {\AA}$^{–1}$) was sampled using a $k_a$ $\times$ $k_b$ $\times$ $k_c$ Pack-Monkhorst k-mesh grid with larger shrinking factors ($a \cdot k_a$ $\geqq$ 60, $b \cdot k_b$ $\geqq$ 60, k$_c$ = 1) compared to those used for geometric optimization ($a \cdot k_a$ $\geqq$ 40, $b \cdot k_b$ $\geqq$ 40, k$_c$ = 1) \cite{monkhorst1976special}.
The cohesive energies were computed using the HSE06-D3 hybrid functional \cite{perdew1992atoms} after geometric optimization with B3LYP-D3. HSE06-D3 incorporates a short-range coulombic operator and has demonstrated accurate predictions for metallic materials \cite{krukau2006influence}. The results of cohesive energies based on different functionals were compared (see Figure S1). It is important to note that the functionals used to calculate each comparative term of binding energy were unified. The calculation of binding energy can be referred to a previous study \cite{nijamudheen2020impact}.

Cohesive energy ($E_{coh}$) refers to the energy needed to evaporate an isolated atom from its bulk metal, and it is calculated using equation~\eqref{eqn:cohesive}.
\begin{equation}\label{eqn:cohesive}
E_{coh} = E_{gas} – \frac{E_{bulk}}{n}
\end{equation}
Here, $n$ represents the number of atoms in the bulk metal, while $E_{gas}$ and $E_{bulk}$ correspond to the total energy of an isolated atom and the bulk metal, respectively.

\begin{figure}[h]
    \centering
    \includegraphics[width=0.47\textwidth]{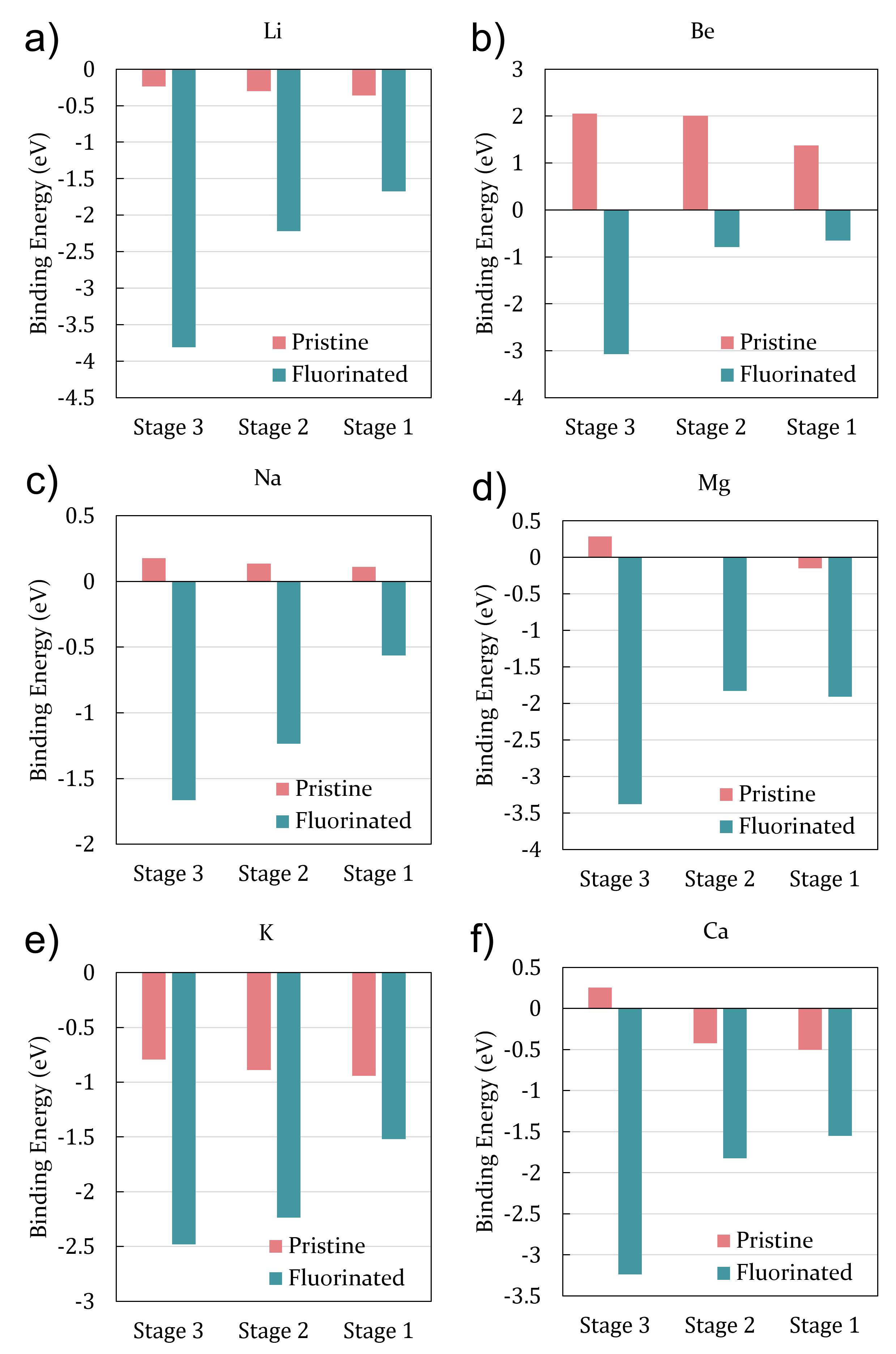}
\caption{\footnotesize Comparison of binding energies between intercalations with and without FSM, covering the cases of (a) Li, (b) Be, (c) Na, (d) Mg, (e) K, and (f) Ca at each stage.}    
    \label{fig:BE_compare}
\end{figure}

\section{Discussion}

The highlights of this study include not only the applicability of FSM to diverse beyond-Li intercalants but also the scheme behind the consequences. The main character and the philosophies around it are illustrated in Figure~\ref{fig:Scheme} first. Prior to demonstrating the effectiveness of the FSM, it is crucial to determine the most thermodynamically stable staging assemblies as a basis for comparison. Since previous studies on staging assemblies were not exhaustive, this work considers all possible configurations. Figure~\ref{fig:BE_pristine} and Figure~\ref{fig:BE_fluorinated} present every assembly at each stage, along with their corresponding binding energies. Specifically, Figure~\ref{fig:BE_pristine} displays the results for pristine 4LG, while Figure~\ref{fig:BE_fluorinated} depicts those for fluorinated 4LG. 
For pristine 4LG at stage 3, structures with the intercalants located at the edge sites are more likely to form stable bindings. At stage 2, assemblies with intercalants positioned both at the edge and middle sites tend to be more energetically favorable. Notably, the intercalation of lithium and potassium within pristine 4LG can be thermodynamically stable even without the assistance of the FSM (Figure~\ref{fig:BE_pristine}). This advantage highlights the widespread usage of LIBs in today's applications and underscores the potential of potassium-ion batteries.
Moving to fluorinated 4LG at stage 3, structures with the intercalant closest to the FSM consistently exhibit the lowest binding energies. Interestingly, as the intercalant moves further away from the FSM, the binding energies gradually increase. This observation suggests that the FSM can significantly impact conditions with low intercalant concentrations, as the preferred host site is just below the FSM, which was not considered in previous work \cite{nijamudheen2020impact}. On the other hand, fluorinated 4LG at stage 2 shows various assemblies for the most stable bindings.

\begin{figure}[h]
    \centering
    \includegraphics[width=0.47\textwidth]{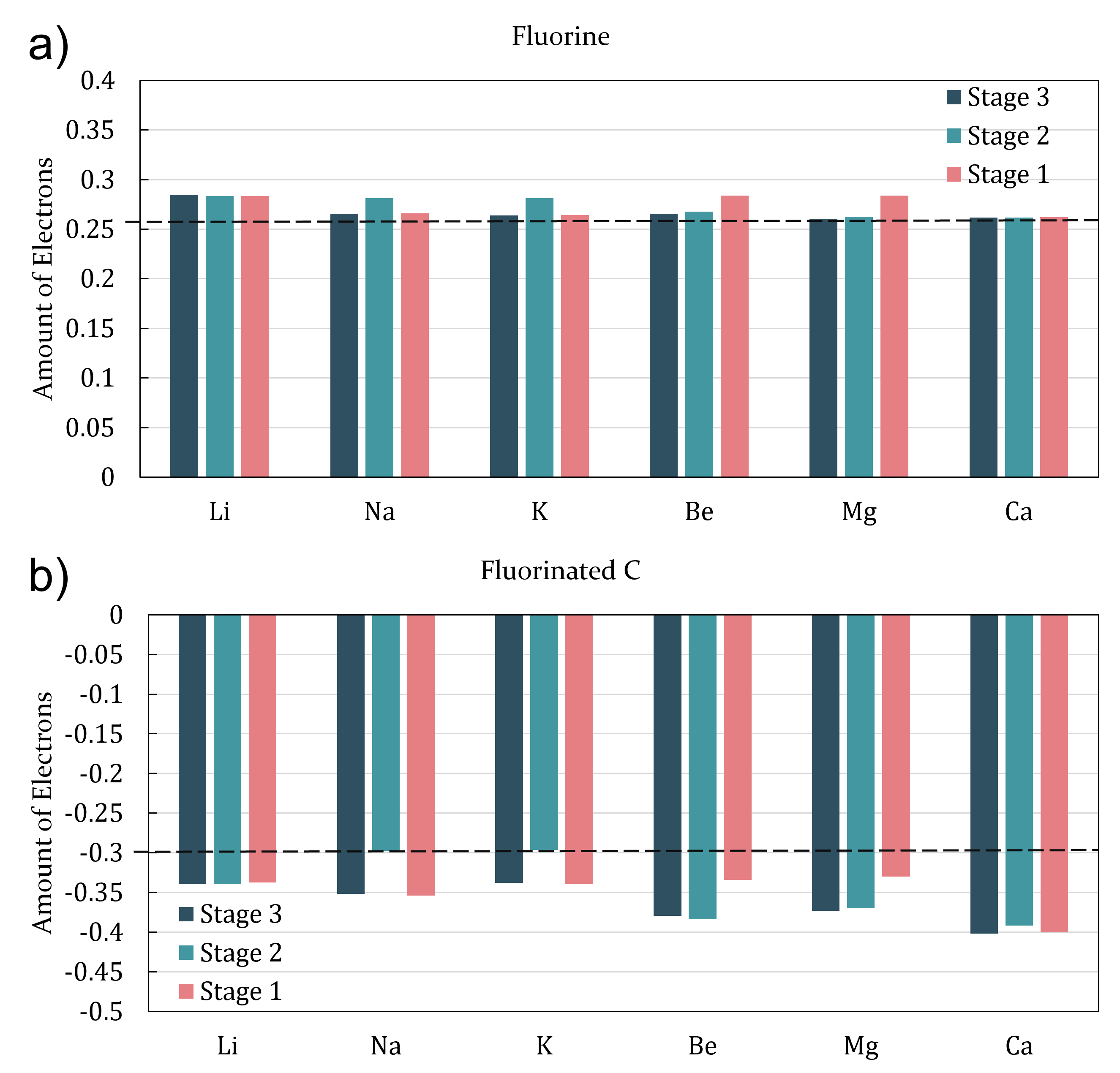}
    \caption{\footnotesize Averaged net flow of electrons for (a) fluorine and (b) fluorinated carbon in the FSM, considering the intercalations of Li, Na, K, Be, Mg, and Ca within fluorinated 4LG at each stage. The dashed lines represent the averaged net flow of electrons in the case without any intercalant.}
    \label{fig:MA_F}
\end{figure}

\begin{figure}[h]
    \centering
    \includegraphics[width=0.47\textwidth]{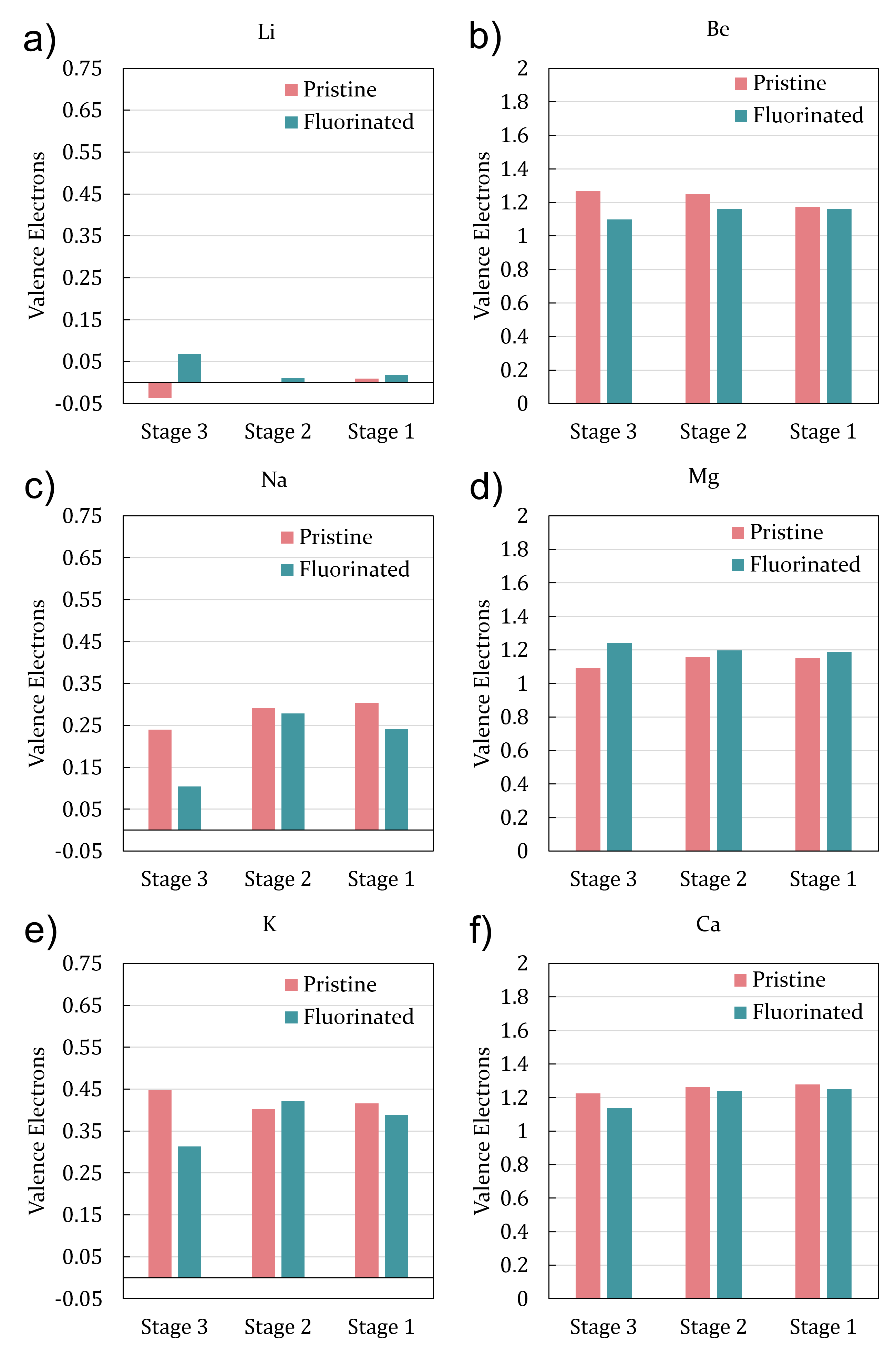}
    \caption{\footnotesize Average valence of intercalants, including (a) Li, (b) Be, (c) Na, (d) Mg, (e) K, and (f) Ca, when intercalated within fluorinated 4LG at each stage.}
    \label{fig:MA_ion}
\end{figure}

With a clear understanding of the dominant assemblies for each element at different stages, a comparison between the intercalation of pristine 4LG and fluorinated 4LG becomes applicable. As shown in Figure~\ref{fig:BE_compare}, the binding energies at each stage decrease significantly with the presence of the FSM, regardless of the element. Remarkably, while Li and K can spontaneously intercalate into pristine 4LG, the intercalation of Na, Be, Mg, and Ca becomes thermodynamically more favorable when fluorinated 4LG is considered. This finding unveils several potential alternatives to LIBs.
Furthermore, the effectiveness of the FSM is particularly prominent at a low ion concentration level, as evidenced by the more pronounced decrease in binding energy at stage 3 compared to stages 2 and 1. As the ion concentration in 4LG increases, from stage 3 to stage 2 or from stage 2 to stage 1, the impact of the FSM weakens. Nevertheless, regardless of the stage, the theoretical specific capacity of fluorinated FLG outperforms pristine graphitic anodes. For instance, as presented in Table S1, the computed capacities of Na (stage 3: 73; stage 2: 137; stage 1: 194 mAhg$^{-1}$) and Ca (stage 3: 139; stage 2: 252; stage 1: 346 mAhg$^{-1}$) ion intercalation in fluorinated 4LG are substantially larger than the experimentally found capacities (Na: 34; Ca: 97 mAhg$^{-1}$) in graphitic electrodes \cite{ge1988electrochemical, park2020stable}. On the other hand, the computed capacity of K-ion intercalation in pristine 4LG at stage 1, which already exhibits favorable binding energy, aligns well with the experimental result ($\sim$ 200 mAhg$^{-1}$) \cite{luo2015potassium}.

\begin{figure*}
\includegraphics[width=0.9\textwidth]{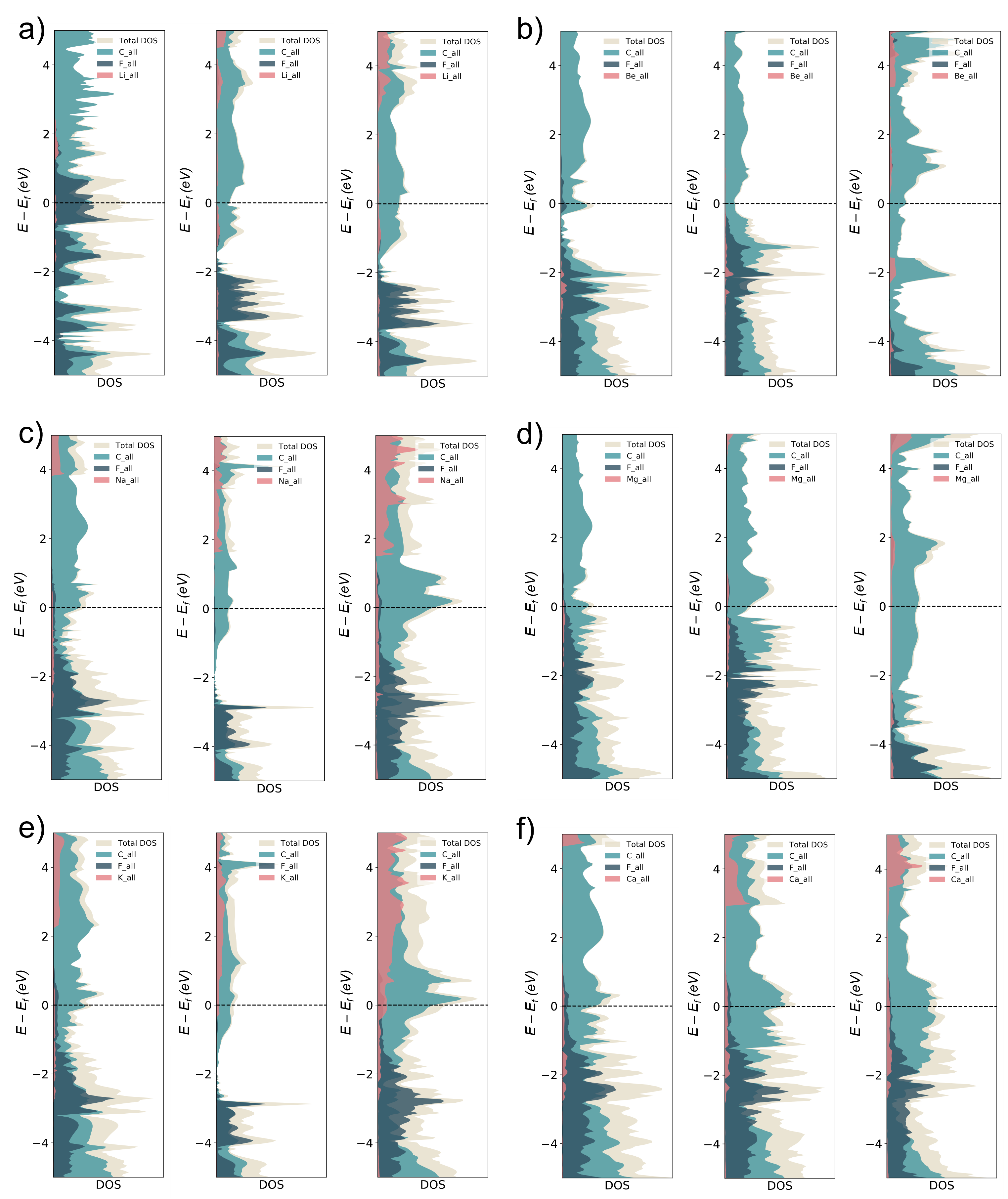}
\centering
\caption{\footnotesize The Partial Density of States (PDOS) of ion intercalations within fluorinated 4LG from stage 3 to stage 1 (from left to right), comprising the cases of (a) Li, (b) Be, (c) Na, (d) Mg, (e) K, and (f) Ca. The PDOS of fluorinated 4LG is shown in Figure S3.}
\label{fig:DOS_F}
\end{figure*}

\begin{figure*}
\includegraphics[width=\textwidth]{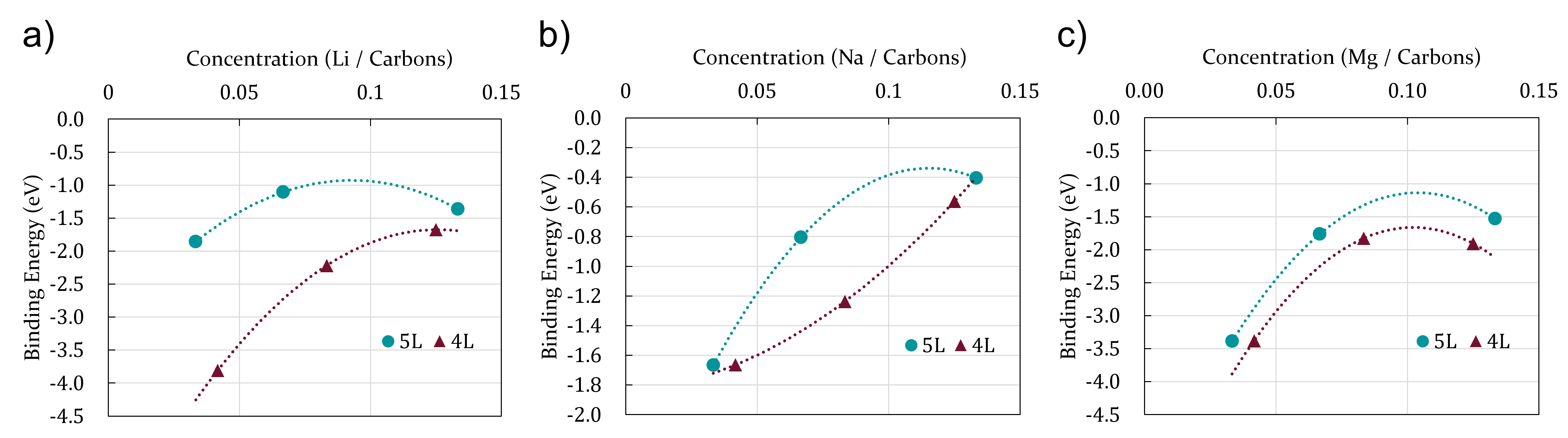}
\centering
\caption{\footnotesize The comparison of binding energies for ion intercalation within fluorinated 4LG and 5LG, encompassing the cases of (a) Li, (b) Na, and (c) Mg at various ion concentrations.}
\label{fig:BE_dimension}
\end{figure*}

This study not only considers intercalant assemblies but also expands the range of alternatives to LIBs. However, the role of FSM in this context remains to be elucidated. Therefore, to deepen our understanding of the underlying mechanism, we investigate the electronic properties. Specifically, we focus on the redistribution of electrons and the electron pathways between each element in the system with FSM. Mulliken analysis was employed to gain insights into the electron distribution. As depicted in Figure~\ref{fig:MA_F}, the fluorine in the FSM gains electrons, while the carbon connected to that fluorine loses electrons. This phenomenon can be attributed to the ultrahigh electronegativity of fluorine, which tends to draw electrons. Moreover, in the presence of intercalants, the fluorine and F-connected carbon show an increased tendency to gain and lose electrons, respectively, compared to fluorinated 4LG without any intercalant. However, the net electron of fluorine and F-connected carbon is negative, leading to the transfer of those missing electrons to other non-fluorinated carbon atoms, making them positively charged, especially the carbons in the FSM layer.
Regarding the intercalants, Figure~\ref{fig:MA_ion} illustrates the averaged valence of intercalants at each stage. The valence of alkali metals ranges from 0 to 0.45, while that of alkaline earth metals ranges from 1 to 1.3. Notably, Li shows limited participation in charge transfer with others in the system, rendering it nearly neutral compared to other types of intercalants. This observation suggests that the intercalation of Li within FLG relies on covalent bonds, while the other intercalants primarily form ionic bonds with negatively charged non-fluorinated carbons. Covalent bonds, held together by electrostatic forces between shared electrons and nuclei, require less energy to form than ionic bonds, partially accounting for the advantage of lithium-ion intercalation.
Interestingly, with FSM, other intercalants at stage 3 tend to retain more electrons (lower valence) compared to higher intercalant concentration stages (stage 2 and stage 1). This behavior aligns with the previous finding that the FSM plays a more effective role at low ion concentration stages, as demonstrated in the comparison of binding energies in Figure~\ref{fig:BE_compare}. Additionally, a larger difference between electronegativities leads to a greater amount of transferred electrons. Hence, compared to either carbon or fluorine, K loses more electrons (higher valence) than Na does and even more than Li does, as shown in Figure~\ref{fig:MA_ion}. However, this effect is less pronounced in alkaline earth metals compared to alkali metals.

The results of Mulliken analysis provide insight into the electron distribution in the system, but the mechanism of electron transfer remains a question. If there is no or insufficient channel for electrons between each element, the flow and drainage of electrons can exhibit distinct behaviors. To shed light on this aspect, electron density of states (DOS) can offer valuable information by indicating the presence of adequate electron channels based on the overlap of DOS. Figure S2 and ~\ref{fig:DOS_F} represent the elemental PDOS, illustrating the DOS of each element.
In the case of pristine 4LG, Figure S2 reveals a clear overlap between carbon and each intercalant, and as the system approaches higher intercalant concentrations from stage 3 to stage 1, this overlap gradually increases. A similar phenomenon is observed in the intercalation of fluorinated 4LG, as shown in Figure~\ref{fig:DOS_F}, indicating the existence of electron channels between carbon and intercalants. However, fluorinated 4LG offers additional channels, not limited to intercalant-carbon interactions, but also open to intercalant-fluorine and carbon-fluorine interactions. This suggests that, with the assistance of the FSM, electron transfer can exhibit greater flexibility, leading to a more stable ground state and lower binding energy. Specifically, the overlap of DOS with FSM around the Fermi energy level, which is more likely to be occupied by electrons, tends to be greater compared to that of pristine 4LG.

Furthermore, the positive impact of FSM on ion intercalation in FLG is not the only contributing factor; the dimension of FLG also plays a significant role. To illustrate this, a comparison of binding energies for Li, Na, and Mg between 4LG and 5LG with FSM is shown in Figure~\ref{fig:BE_dimension}. Given that the configurations of each stage differ between 4LG and 5LG, the intercalant concentration was considered instead of the stage. Similar to 4LG, FSM in 5LG proves to be more effective at lower ion concentrations and in staging assemblies with intercalants positioned close to it, as depicted in Figure~\ref{fig:BE_dimension} and Figure S4, respectively. Additionally, the binding energy in 5LG gradually increases as the intercalant concentration rises. However, it is noteworthy that the intercalation within fluorinated 4LG outperforms that of fluorinated 5LG, as it requires less energy for binding at each stage. This finding provides a valuable clue for practical energy storage applications, suggesting that a higher degree of graphitic exfoliation for graphene could further enhance the capacity of FLG electrodes.

\section{Conclusion}

In conclusion, this comprehensive study thoroughly explores the application of fluorinated FLG in diverse ion intercalations (Li, Na, K, Be, Mg, and Ca) by considering staging assemblies, electronic properties, and dimensions. By comparing the binding energies of staging configurations, we found that FSM proves to be remarkably effective in facilitating intercalation at low ion concentrations. Notably, the most preferable host layer at stage 3 is located beneath FSM, and as the intercalant moves farther away from FSM, the binding energy gradually increases. This trend applies to all six types of ions. Most importantly, FSM was observed to either reverse nonspontaneous intercalations (Na, Be, Mg, and Ca) or further facilitate spontaneous interactions (Li and K) at each stage, opening up possibilities for multiple potential alternatives to LIBs.
Moreover, our electronic analysis sheds light on the mechanism of FSM-assisted intercalation. Specifically, the Mulliken analysis suggests that Li-ion intercalation may involve covalent bonds, while other intercalants tend to undergo ionic interactions with significant charge transfer. Additionally, based on the PDOS, FSM creates additional channels for flexible electron transfer, enabling the system to reach a more stable ground state.
Furthermore, we found that the number of layers can influence the theoretical capacity of FLG. Intercalation in fluorinated 4LG was observed to be more thermodynamically stable than in fluorinated 5LG. In summary, this work not only highlights the advantages of applying FSM in beyond-lithium-ion batteries but also provides dimensional and electronic perspectives on electrode engineering. These findings contribute valuable insights to the development of advanced energy storage systems. 

\bibliography{main}

\clearpage
\newpage

\onecolumngrid

\section{Supporting Information}

\renewcommand{\figurename}{Figure S\!\!}
\renewcommand{\tablename}{Table S\!\!}
\setcounter{figure}{0}  


\begin{table}[h]
\centering
\caption{The maximum theoretical specific capacity of each intercalated system.}
\label{table:capacity}
\begin{tabular}{lccc} \hline
\textbf{Model}  & \textbf{Stage 3 / $mAh~g^{-1}$} & \textbf{Stage 2 /$mAh~g^{-1}$} & \textbf{Stage 1 $mAh~g^{-1}$} \\ \hline
P~4LG-Li & 90.82      & 177.46     & 260.22 \\
F~4LG-Li & 76.12      & 149.30     & 219.71 \\
P~4LG-Na & 86.14$^*$  & 160.42$^*$ & 225.15$^*$ \\
F~4LG-Na & 72.80      & 137.05     & 194.17 \\
P~4LG-K  & 81.90      & 146.32     & 198.32 \\
F~4LG-K  & 69.75      & 126.63     & 173.88 \\
P~4LG-Be & 180.37$^*$ & 350.13$^*$ & 510.18$^*$ \\
F~4LG-Be & 151.35     & 295.20     & 432.08 \\
P~4LG-Mg & 171.55$^*$ & 318.34$^*$ & 445.37 \\
F~4LG-Mg & 145.09     & 272.27     & 384.67 \\
P~4LG-Ca & 163.31$^*$ & 291.08     & 393.78 \\
F~4LG-Ca & 139.15     & 252.08     & 345.57 \\
F~5LG-Li & 63.19      & 124.35     & 240.94 \\
F~5LG-Na & 60.80      & 115.73     & 210.57 \\
F~5LG-Mg & 121.41     & 230.16     & 416.84 \\ \hline 
\multicolumn{4}{l}{\small *Notes:} \\
\multicolumn{4}{l}{\small 1.~$^*$ represents the system with thermodynamically unfavorable binding energy.} \\
\multicolumn{4}{l}{\small 2.~P stands for pristine, while F means fluorinated.} \\
\end{tabular}

\end{table}

\begin{figure}[h]
    \centering
    \includegraphics[width=0.7\textwidth]{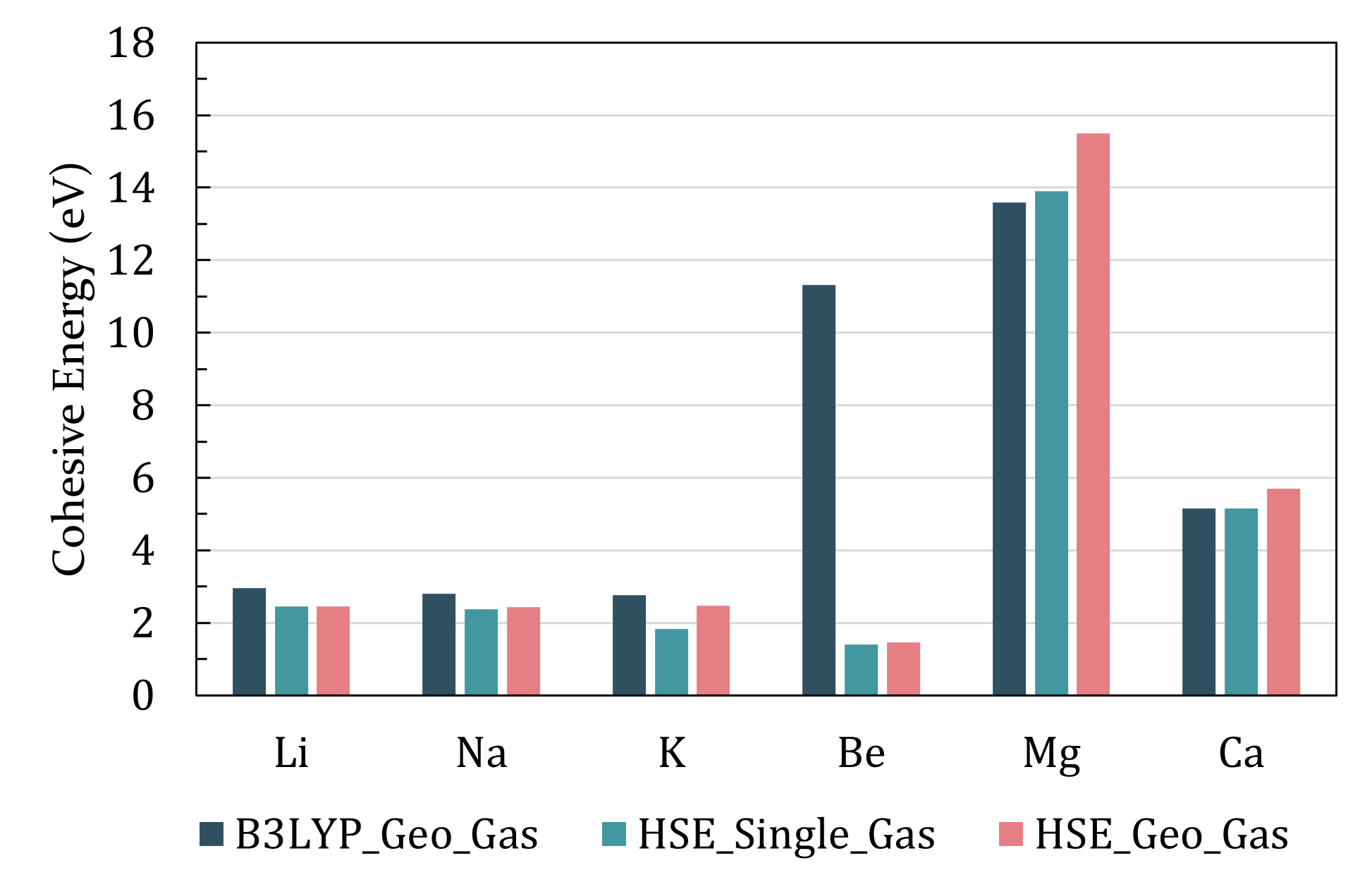}
    \caption{The comparison of cohesive energy calculated in different functionals, including B3LYP-based geometric optimization, B3LYP-based geometric optimization with HES06-based singlepoint calculation, and  HES06-based geometric optimization.}
    \label{fig:Cohesive}
\end{figure}

\begin{figure}[h]
    \centering
    \includegraphics[width=0.9\textwidth]{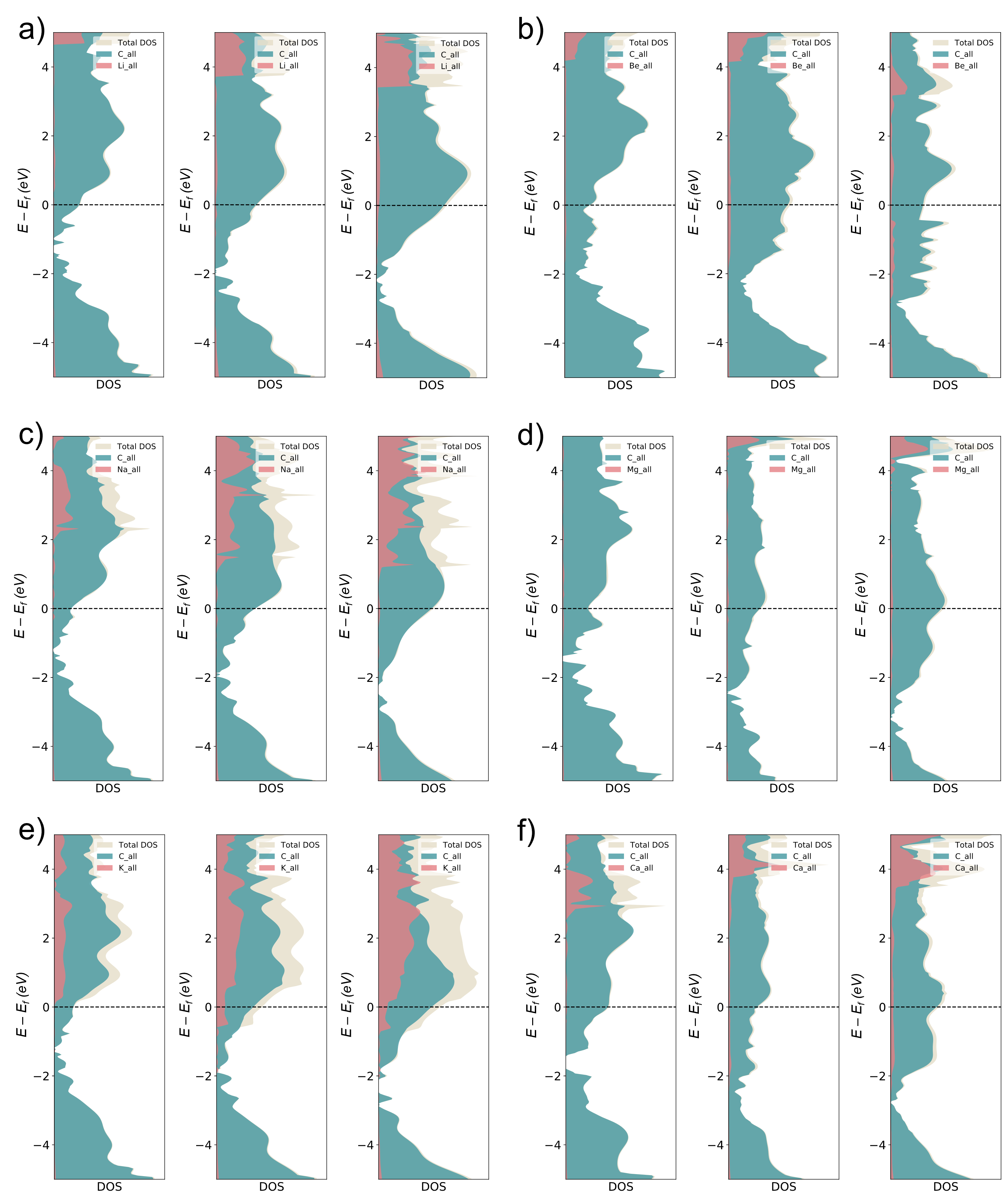}
    \caption{\footnotesize The Partial Density of States (PDOS) of ion intercalations within pristine 4LG from stage 3 to stage 1 (from left to right), comprising the cases of (a) Li, (b) Be, (c) Na, (d) Mg, (e) K, and (f) Ca.}
    \label{fig:DOS_P}
\end{figure}

\begin{figure}[h]
    \centering
    \includegraphics[width=0.25\textwidth]{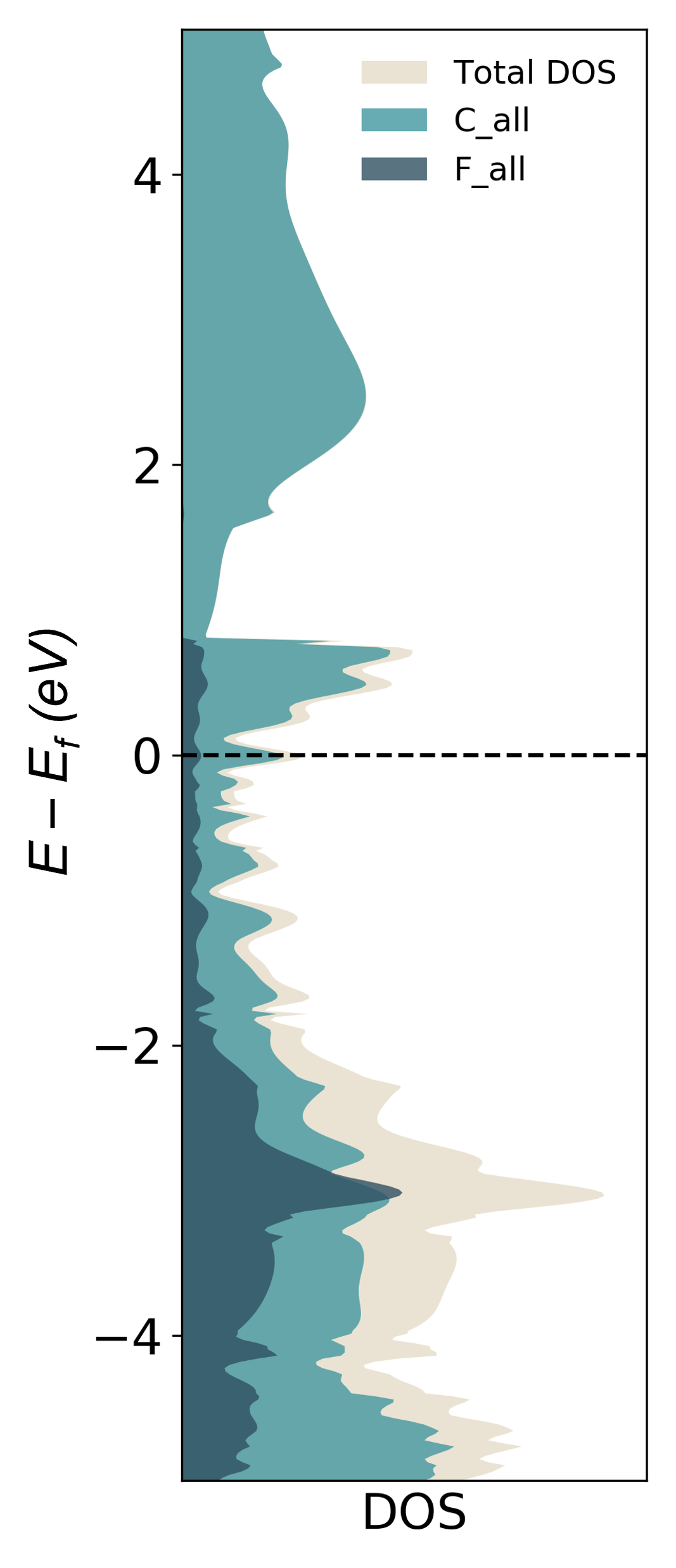}
    \caption{\centering The PDOS of fluorinated 4LG.}
    \label{fig:DOS_FSM}
\end{figure}

\begin{figure}[h]
    \centering
    \includegraphics[width=0.8\textwidth]{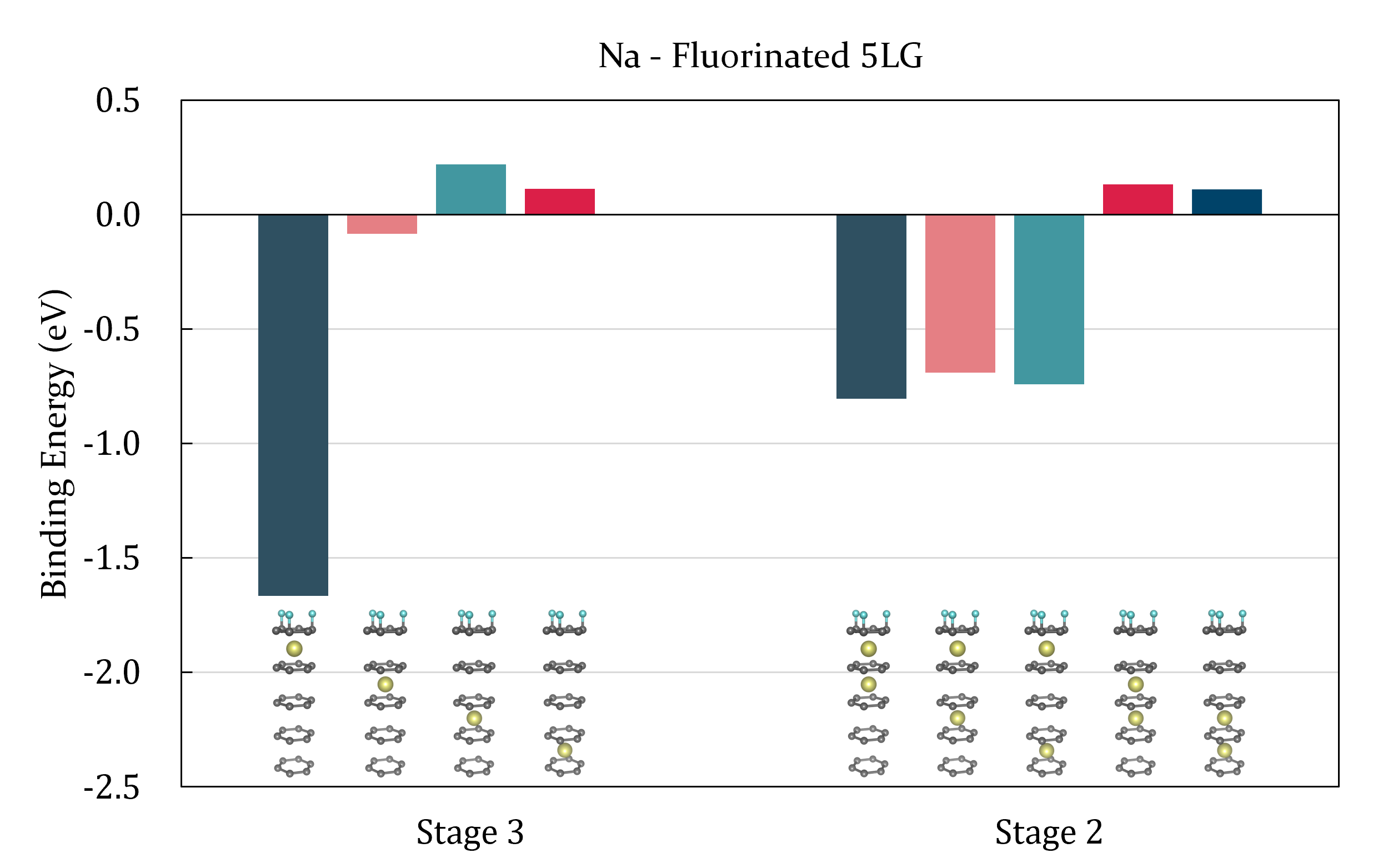}
    \caption{The comparison of binding energies of each staging assembles at stage 3 (left) and stage 2 (right) in the case of Na ion intercalation within fluorinated 5LG. The inserted schematic diagrams are aligned with the corresponding binding energies.}
    \label{fig:5LG_Na}
\end{figure}

\end{document}